\documentclass[a4paper,twoside]{article}

\usepackage{epsfig}
\usepackage{subcaption}
\usepackage{calc}
\usepackage{amssymb}
\usepackage{amstext}
\usepackage{amsmath}
\usepackage{amsthm}
\usepackage{multicol}
\usepackage{pslatex}
\usepackage{apalike}
\usepackage{algorithm2e}
\usepackage[bottom]{footmisc}
\usepackage{SCITEPRESS}    
\usepackage{url} 
\usepackage{booktabs}

\usepackage{tikz}
\usetikzlibrary{shapes, arrows, positioning, fit, backgrounds, calc}
 
\begin{document}

\title{Privacy-Preserving Compliance on Public Ledgers via Selective Disclosure Authorization Schemes}


\author{\authorname{Supriya Khadka\orcidAuthor{0009-0008-4126-624X} and Sanchari Das\orcidAuthor{0000-0003-1299-7867}}
\affiliation{George Mason University, Fairfax, Virginia, USA}
\email{\{skhadk, sdas35\}@gmu.edu}
}


\abstract{Public distributed ledgers enforce integrity through radical transparency, creating tension with data minimization principles required for regulatory compliance. While Zero-Knowledge Proofs (ZKPs) offer a theoretical privacy solution, existing constructions often overlook adversarial constraints in smart contract environments. Specifically, the asynchronous decoupling of off-chain proof generation from on-chain submission introduces front-running and proof-reuse risks in public mempools. In this work, we formalize \emph{Selective Disclosure Authorization Schemes (SDAS)}, a cryptographic primitive for granular and revocable compliance checks on public ledgers without revealing the underlying witness. We define a security model for SDAS, introducing \emph{Ledger-Bound Attribute Unlinkability} and \emph{Context-Aware Sender Binding} to capture how valid proofs remain bound to their intended authorization context. To validate sender binding, we present \emph{ZK-Compliance}, an Ethereum-based instantiation that operationalizes a user-controlled ``Grant, Verify, Revoke'' lifecycle. We implement the sender-binding component using a 14-constraint Circom circuit that anchors the zero-knowledge proof to the executing on-chain sender address. Our Sepolia evaluation confirms practical viability: browser-based proof generation executes in under $200$ ms, and on-chain verification costs $240,512$ gas, neutralizing proof reuse by different callers while preserving strict attribute privacy.
}

\onecolumn \maketitle \normalsize \setcounter{footnote}{0} \vfill

\section{\uppercase{Introduction}}
\label{sec:introduction}
Public distributed ledgers achieve trustless consensus through radical transparency, allowing participants to audit global state without centralized intermediaries. Systems such as Ethereum preserve integrity by recording state transitions on an immutable ledger~\cite{conti2018survey,kishnani2023blockchain,agarwal2025systematic,kishnani2025security}. While this supports auditability and censorship resistance, it creates an adversarial environment for privacy~\cite{buterin2016privacy,khadka2026grant,tazi2022sok}. Public verifiability conflicts with data minimization, which requires systems to expose only information necessary for a specific purpose.

This tension is especially acute for regulatory compliance and identity verification. Decentralized applications (dApps) increasingly need to enforce eligibility predicates, such as age thresholds or residency requirements, before authorizing state transitions~\cite{belchior2021survey}. Posting identity attributes on-chain forces users into a ``transparency trap'' that permanently links real-world identity information to transactional history~\cite{buterin2016privacy,khadka2026poster}. On public ledgers, such over-disclosure is globally visible, irreversible, and difficult to reason about ex ante~\cite{Bernabe2019}.

In practice, many blockchain applications address regulatory pressure by outsourcing identity verification to trusted third parties, centralized exchanges, or compliance providers~\cite{belchior2021survey}. Although this satisfies compliance requirements, it reintroduces centralized identity repositories that are attractive targets for compromise~\cite{xian2025survey,zamani2020security}. Centralized verification also expands the attack surface through credential exfiltration, unauthorized data sharing, and retention of sensitive attributes~\cite{almasi2020protecting,tazi2024sok}. More fundamentally, these models provide limited support for granular revocation or time-bounded access, leaving users with little control over how long or for what purpose their attributes are used~\cite{bruhner2023bridging,Ishmaev2021}.

Zero-Knowledge Proofs (ZKPs), particularly zk-SNARKs~\cite{ben2014succinct,baghery2020simulation}, offer a promising alternative by allowing a prover to verify a statement without revealing the underlying witness~\cite{goldwasser1985zkp}. However, applying ZKPs to smart contract authorization introduces execution-layer challenges that standard knowledge soundness does not address. In a browser-based dApp, a proof is generated locally at time $t$ but submitted through a public mempool for inclusion at time $t+\delta$. This delay creates a window in which an adversary can observe a valid proof and replay it in another transaction context~\cite{daian2020flashboys}. Standard SNARKs guarantee witness knowledge, but they do not inherently guarantee that the prover is also the sender of the on-chain transaction.

In this work, we address this gap by formalizing the \textbf{Selective Disclosure Authorization Scheme (SDAS)}. Unlike general-purpose transactional privacy protocols that obfuscate the transaction graph~\cite{sasson2014zerocash,Bernabe2019}, SDAS focuses on predicate-level authorization. We instantiate this framework through \textbf{ZK-Compliance}, an Ethereum-based system that operationalizes a user-controlled ``Grant, Verify, Revoke'' lifecycle and implements a sender-binding component against mempool front-running. By treating authorization as a verifiable state transition rather than a stateless proof presentation, ZK-Compliance reduces reliance on persistent identity custodians while preserving attribute privacy. Our core contributions are:

\begin{itemize}
    \item We introduce SDAS, a public-ledger authorization architecture that separates \textit{attribute validity} from \textit{stateful on-chain authorization} and models authorization as a verifiable state transition.

    \item We define \textit{Context-Aware Sender Binding} and \textit{Ledger-Bound Attribute Unlinkability}, and instantiate the sender-binding component in ZK-Compliance, a browser-native Ethereum prototype using a $14$-constraint Circom circuit that binds the proof to the executing \texttt{msg.sender}.
    
    \item We evaluate ZK-Compliance on Sepolia, showing that client-side proof generation completes in under $200$ ms and on-chain verification costs $240,512$ gas, demonstrating the feasibility of sender-bound selective disclosure authorization.

\end{itemize}

\section{\uppercase{Related Work}}

\subsection{Predicate-Level Blockchain Privacy}
Early blockchain privacy research focused on protecting transactional metadata. Systems such as Zerocash~\cite{sasson2014zerocash} use zk-SNARKs~\cite{ben2014succinct} to hide transaction senders, recipients, and values, while later constructions such as Groth16~\cite{groth2016size} improve succinctness and verification efficiency. These systems provide strong confidentiality for value transfer and fungibility, but operate mainly at the transaction-graph level rather than on compliance-oriented identity predicates. Surveys further show that integrating such privacy protocols into general-purpose decentralized applications remains difficult without affecting composability or interoperability~\cite{zhang2019security,Bernabe2019}. In contrast, our work applies zk-SNARKs to predicate-level authorization, allowing users to prove attributes such as age or eligibility while remaining compatible with standard Ethereum smart contracts.

Selective disclosure is closely related to anonymous credentials, which allow users to prove predicates over attributes without revealing their identity~\cite{camenisch2001efficient}. However, classical anonymous credential models often assume secure channels and fresh pseudonyms for unlinkability. This assumption does not directly transfer to account-based blockchains, where the network is a public broadcast channel and gas payments can link pseudonymous addresses to funding sources~\cite{bunz2020zether,bowe2020zexe}. Our model therefore distinguishes transactional linkability from attribute linkability: the ledger may reveal the executing address, but the eligibility witness remains hidden.

\subsection{Revocable Decentralized Identity}
Decentralized identity research has converged around Self-Sovereign Identity (SSI) and the W3C Verifiable Credentials standard~\cite{w3c2022vc}. These systems aim to reduce reliance on centralized identity providers and support selective disclosure. In practice, however, SSI deployments often depend on a multi-layer trust stack involving DID methods, credential registries, specialized wallets, and issuer-managed revocation mechanisms~\cite{Krul_2024,Ishmaev2021}. Prior work also shows that users often lack meaningful mechanisms to enforce purpose limitation or revoke access once credentials have been presented~\cite{bruhner2023bridging}.

Existing ZK identity frameworks such as Polygon ID, Holonym, and OutDID share the goal of privacy-preserving verification~\cite{polygonid_blog2022,holonym_docs,outdid_homepage}. However, they mainly focus on credential issuance, proof of personhood, credential portability, or hardware-backed identity verification. For example, issuer-centric revocation mechanisms such as revocation trees can invalidate credentials, but they do not necessarily give users direct control over a specific dApp's on-chain authorization state~\cite{privado_revocation}. Some systems also rely on specialized mobile wallets, NFC-enabled identity documents, or persistent identifiers, which can introduce friction or linkability concerns~\cite{sbt_seaflux}. Our approach instead focuses on the authorization lifecycle after eligibility has been proven: access is granted, checked, and revoked through smart-contract state while the underlying identity attribute remains private.

\subsection{Context-Bound Authorization}
Smart contract access control commonly uses access-control lists or role-based access control, including OpenZeppelin's \texttt{AccessControl}~\cite{openzeppelin_access_control,cruz2018rbac,podapati2025sok}. These mechanisms are simple and widely used, but bind permissions directly to public addresses. As a result, proving eligibility can permanently associate authorization status with an on-chain identity. Static authorization is also vulnerable to confused-deputy-style failures in composable smart contract environments~\cite{gritti2023confusum,ajayi2024enhancing}.

Our work is closer to capability-based access control, where access is granted through possession of an authorization object rather than repeated identity inspection~\cite{miller2006robust}. However, stateless or transferable capabilities are problematic in public mempool environments: once a proof is visible, an adversary may attempt to replay or front-run it from another transaction context. SDAS addresses this by treating the ZK proof as a non-transferable, context-bound capability~\cite{khadka2026towards,podapati2025sok}. In our prototype, the proof is bound to the executing sender address, so a proof observed in the mempool cannot be reused by a different caller. This design differs from approaches based on trusted hardware or attribute-based encryption, which introduce additional hardware trust assumptions or computational overhead~\cite{zyskind2015enigmadecentralizedcomputationplatform,Cheng_2019,zhang2019security}. Instead, ZK-Compliance relies on lightweight zk-SNARK verification and explicit execution-context binding to enforce privacy-preserving authorization on-chain.

\section{\uppercase{Formal Model: SDAS}}

In this section, we formalize the Selective Disclosure Authorization Scheme (SDAS) and define the adversarial environment in which it operates. 

\subsection{The Execution Context}
To mitigate replay and ambiguity threats, the SDAS framework treats the blockchain not merely as a ledger, but as a global state transition machine operating over specific execution contexts. For every transaction execution, the machine provides an immutable context tuple $\Gamma$. We standardize this tuple as:

$$ \Gamma := (\chi_{addr}, \chi_{chain}, \chi_{contract}) $$

This corresponds to the Ethereum Virtual Machine (EVM) variables \texttt{msg.sender}, \texttt{block.chainid}, and \texttt{address(this)}. We assume the underlying consensus layer guarantees the integrity of $\Gamma$, with the sender authenticated via ECDSA and the chain ID fixed by protocol configuration. A valid authorization proof must attest to a relation $\mathcal{R}(w, \Gamma)$, ensuring it is bound to this coordinate space; our prototype instantiates this binding for \texttt{msg.sender}. Here, $w$ denotes the private witness, comprising the user's secret attributes and auxiliary randomness for proof generation.


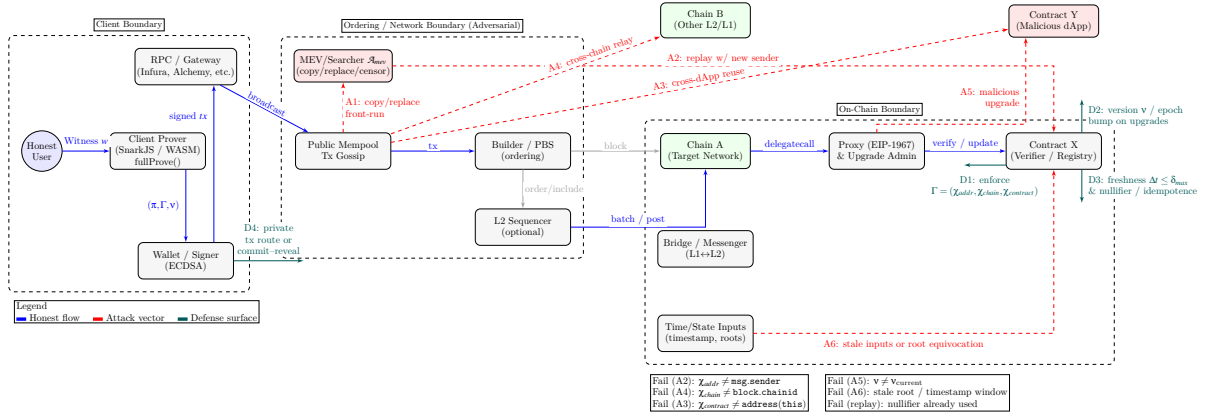
\begin{figure*}[!t]
\centering
\resizebox{1\linewidth}{!}{%
\begin{tikzpicture}[
scale=0.78, transform shape,
font=\normalsize,
node distance=1.6cm,
>=latex',
every node/.append style={fill=white, inner sep=2pt}
]
\tikzstyle{actor}=[circle, draw, fill=blue!10, minimum size=1.2cm, align=center]
\tikzstyle{component}=[rectangle, draw, rounded corners, fill=gray!7, minimum width=3.4cm, minimum height=1.3cm, align=center]
\tikzstyle{chain}=[rectangle, draw, rounded corners, fill=green!7, minimum width=3.2cm, minimum height=1.25cm, align=center]
\tikzstyle{adv}=[rectangle, draw, rounded corners, fill=red!10, minimum width=3.2cm, minimum height=1.15cm, align=center]
\tikzstyle{note}=[rectangle, draw, fill=white, align=left, font=\footnotesize]
\tikzstyle{boundary}=[draw, dashed, rounded corners, inner sep=10pt, fill=none]
\tikzstyle{honest}=[->, thick, blue]
\tikzstyle{attack}=[->, thick, red, dashed]
\tikzstyle{defend}=[->, thick, teal!70!black]
\tikzstyle{grayflow}=[->, thick, gray!70]

\node[actor] (user) {Honest\\User};
\node[component, right=of user, xshift=0.1cm] (client) {Client Prover\\(SnarkJS / WASM)\\\texttt{fullProve()}};

\node[component, below=of client, yshift=-1.0cm, xshift=1cm] (wallet) {Wallet / Signer\\(ECDSA)};

\node[component, above=of client, yshift=0.1cm, xshift=1cm] (rpc) {RPC / Gateway\\(Infura, Alchemy, etc.)};

\node[component, right=of client, xshift=1.6cm] (mempool) {Public Mempool\\Tx Gossip};
\node[adv, above=of mempool, yshift=0.2cm] (mev) {MEV/Searcher $\mathcal{A}_{mev}$\\(copy/replace/censor)};
\node[component, right=of mempool, xshift=1.4cm] (builder) {Builder / PBS\\(ordering)};
\node[component, below=of builder, yshift=0.2cm] (sequencer) {L2 Sequencer\\(optional)};
\node[chain, right=of builder, xshift=1.6cm] (chainA) {Chain A\\(Target Network)};
\node[chain, above=of chainA, yshift=1.8cm] (chainB) {Chain B\\(Other L2/L1)};
\node[component, right=of chainA, xshift=1.2cm] (proxy) {Proxy (EIP-1967)\\\& Upgrade Admin};
\node[component, right=of proxy, xshift=1.3cm] (targetX) {Contract X\\(Verifier / Registry)};
\node[adv, above=of targetX, yshift=1.8cm] (malY) {Contract Y\\(Malicious dApp)};
\node[component, below=of chainA, yshift=-0.6cm] (bridge) {Bridge / Messenger\\(L1$\leftrightarrow$L2)};
\node[component, below=of bridge, yshift=-0.1cm] (oracle) {Time/State Inputs\\(timestamp, roots)};

\node[boundary, fit=(user) (client) (wallet) (rpc)] (tb_client) {};
\node[boundary, fit=(mempool) (builder) (sequencer) (mev)] (tb_net) {};
\node[boundary, fit=(chainA) (proxy) (targetX) (bridge) (oracle)] (tb_chain) {};

\node[note, anchor=south] at ($(tb_client.north)+(0,0.15cm)$) {\textbf{Client Boundary}};
\node[note, anchor=south] at ($(tb_net.north)+(0,0.15cm)$) {\textbf{Ordering / Network Boundary (Adversarial)}};
\node[note, anchor=south] at ($(tb_chain.north)+(0,0.15cm)$) {\textbf{On-Chain Boundary}};

\draw[honest] (user) -- node[midway, above=0.15cm] {Witness $w$} (client);
\draw[honest] (wallet.north |- client.south) -- node[midway, left=0.15cm] {$(\pi,\Gamma,\nu)$} (wallet.north);

\draw[honest] ([xshift=1cm]wallet.north) -- node[pos=0.8, left=0.15] {signed $tx$} ([xshift=1cm]rpc.south);
\draw[honest] (rpc) -- node[midway, sloped, above] {broadcast} (mempool);
\draw[honest] (mempool) -- node[midway, above] {tx} (builder);

\draw[grayflow] (builder) -- node[midway, right] {order/include} (sequencer);
\draw[honest] (sequencer) -| node[pos=0.25, above, align=center] {batch / post} (chainA);
\draw[grayflow] (builder) -- node[midway, above] {block} (chainA);

\draw[honest] (chainA) -- node[midway, above] {delegatecall} (proxy);
\draw[honest] (proxy) -- node[midway, above] {verify / update} (targetX);

\draw[attack] (mempool) -- node[midway, right, align=left] {A1: copy/replace\\front-run} (mev);
\draw[attack] (mev) -| node[pos=0.25, above, align=left] {A2: replay w/ new sender} (targetX);
\draw[attack] (mempool) -- node[midway, sloped, above, align=left] {A3: cross-dApp reuse} (malY);
\draw[attack] (mempool) -- node[pos=0.75, sloped, above, align=left] {A4: cross-chain relay} (chainB);
\draw[attack] (proxy.north) -- ++(0,0.3cm) -| 
    node[pos=0.05, xshift=3.5cm, yshift=0.5cm, above, align=right] {A5: malicious \\ upgrade} 
    ([xshift=-1cm]malY.south);
    
\draw[attack] (oracle) -| node[pos=0.25, below=0.15cm, align=left] {A6: stale inputs or root equivocation} (targetX);

\draw[defend] ([yshift=-0.5cm]targetX.west) -- ++(-1.5cm,0) 
    node[midway, above, align=center, yshift=-1.2cm] {D1: enforce\\$\Gamma=(\chi_{addr},\chi_{chain},\chi_{contract})$};

\draw[defend] ([xshift=1cm]targetX.north) -- ++(0,1.2cm) 
    node[midway, right=0.15cm, align=left] {D2: version $\nu$ / epoch\\bump on upgrades};

\draw[defend] ([xshift=1cm]targetX.south) -- ++(0,-1.2cm) 
    node[midway, right=0.15cm, align=left] {D3: freshness $\Delta t\le \delta_{max}$\\\& nullifier / idempotence};
    
\draw[defend] (wallet.east) -- ++(2.5cm,0) node[midway, above=0.15cm, align=center] {D4: private\\ tx route or \\commit--reveal};

\node[note, anchor=north west, font=\normalsize] (failA)
at ($(tb_chain.south west)+(0.2cm,-0.35cm)$)
{\textbf{Fail (A2)}: $\chi_{addr} \neq \mathtt{msg.sender}$\\
\textbf{Fail (A4)}: $\chi_{chain} \neq \mathtt{block.chainid}$\\
\textbf{Fail (A3)}: $\chi_{contract} \neq \mathtt{address(this)}$};

\node[note, anchor=north west, font=\normalsize] (failB)
at ($(failA.north east)+(0.6cm,0)$)
{\textbf{Fail (A5)}: $\nu \neq \nu_{\text{current}}$\\
\textbf{Fail (A6)}: stale root / timestamp window\\
\textbf{Fail (replay)}: nullifier already used};

\node[note, anchor=north west, font=\normalsize]
at ($(tb_client.south west)+(0.2cm,-0.3cm)$)
{\textbf{Legend}\\
\textcolor{blue}{\rule{0.35cm}{0.12cm}} Honest flow \quad
\textcolor{red}{\rule{0.35cm}{0.12cm}} Attack vector \quad
\textcolor{teal!70!black}{\rule{0.35cm}{0.12cm}} Defense surface};

\end{tikzpicture}
}
\caption{ZK-Compliance architecture and threat model for preventing cross-sender proof reuse.\vspace{-3mm}}
\label{fig:threat_model}
\end{figure*}

\subsection{Threat Model and Assumptions}
We consider a powerful active adversary $\mathcal{A}$ in a permissionless blockchain environment. Unlike classical secure channels, the network layer is public. The adversary has the following capabilities:
\begin{itemize}
    \item \textbf{Mempool Observation:} $\mathcal{A}$ has full read access to the P2P network layer and can observe pending transactions before they are included in a block.
    \item \textbf{Cross-Domain Replay:} $\mathcal{A}$ can relay observed proofs to other compatible blockchains (e.g., Ethereum Mainnet to Arbitrum) or other dApps to attempt unauthorized access.
    \item \textbf{Malicious Prover:} $\mathcal{A}$ can generate proofs for arbitrary inputs if they possess the witness data.
\end{itemize}

We visualize the threat model in Figure~\ref{fig:threat_model} and model the following operational threat vectors:
\begin{itemize}
    \item \textbf{Mempool Front-Running / Proof Reuse by Different Sender:} An adversary observing a valid transaction $tx_{hon}$ may attempt to clone the proof and submit an identical transaction $tx_{adv}$ with a higher gas price. In the full SDAS model, this is mitigated by enforcing $\Gamma$ in the public input vector. The prototype implements the sender component, while chain and contract binding remain extensions.
    \item \textbf{Cross-Chain and Cross-Contract Replay:} A proof valid for Service A on Chain 1 theoretically satisfies the arithmetic constraints of Service B on Chain 2. We mitigate this by enforcing $\Gamma$ in the public input vector. Any mismatch in the execution environment invalidates the proof.
    \item \textbf{Timestamp Manipulation and Stale Proofs:} Miners can manipulate \texttt{block.timestamp} within a narrow tolerance. More critically, mempool latency can delay transaction inclusion. To prevent the replay of stale proofs, the framework enforces a Proof Freshness Window (e.g., 900 seconds) checked against the block timestamp.
\end{itemize}

\subsection{Syntax of SDAS}
An SDAS is a tuple of polynomial-time algorithms $\Pi_{SDAS} = (\text{Setup}, \text{KeyGen}, \text{ProveAuth}, \text{VerifyAuth}, \text{Revoke})$ defined as follows:

\begin{itemize}
    \item $pp \leftarrow \text{Setup}(1^{\lambda}, \mathcal{R})$: Run by a protocol instantiation. It takes a security parameter $\lambda$ and a compliance relation $\mathcal{R}$ to output public parameters $pp$, including proving and verification keys, and an initial state root $rt$.
    \item $(sk_{id}, pk_{id}) \leftarrow \text{KeyGen}(pp)$: Run by the User. Outputs a secret identity witness $sk_{id}$ and a public identity commitment $pk_{id}$. 
    \item $\pi_{auth} \leftarrow \text{ProveAuth}(pp, sk_{id}, rt, \Gamma_{target})$: Run by the User. It takes the secret attributes, the valid set root, and the target execution context $\Gamma_{target}$. It outputs an authorization proof $\pi_{auth}$ cryptographically bound to $\Gamma_{target}$.
    \item $b \leftarrow \text{VerifyAuth}(pp, \pi_{auth}, rt, \Gamma_{curr}, \sigma_{t})$: Run by the Smart Contract. Takes the proof, the current environment $\Gamma_{curr}$, and ledger state $\sigma_{t}$. Outputs 1 (Authorized) only if the proof is valid relative to $rt$ and $\Gamma_{target} == \Gamma_{curr}$, updating the state to $\sigma_{t+1}$.
    \item $\sigma_{t+1} \leftarrow \text{Revoke}(\chi_{addr}, \sigma_{t})$: Run by the User. Updates the global state to explicitly invalidate the access record associated with $\chi_{addr}$.
\end{itemize}

\subsection{Security Definitions}
We define security of an SDAS through two cryptographic properties modeled via games between a challenger $\mathcal{C}$ and an adversary $\mathcal{A}$.

\textbf{Property 1: Ledger-Bound Attribute Unlinkability.} We distinguish between transactional linkability (inherent to the account-based model) and attribute linkability. This property ensures that an on-chain authorization record reveals no information about the specific underlying witness. 

In Game $\text{Unlink}_{\mathcal{A}, \Pi}(\lambda)$, $\mathcal{A}$ chooses a target context $\Gamma^{*}$ and two distinct valid witnesses $(sk_{0}, sk_{1})$. $\mathcal{C}$ flips a coin $b \in \{0,1\}$ and generates a challenge proof using $sk_{b}$. $\mathcal{A}$ wins if they correctly guess $b$. We say the scheme has Attribute Unlinkability if $|\Pr[\mathcal{A} \text{ wins}] - 1/2| \le \text{negl}(\lambda)$.

\textbf{Property 2: Context-Aware Binding and Replay Resistance.} This property ensures a proof generated for a specific execution context cannot be used in any other context, nor replayed within the same context to force redundant state execution. 

In Game $\text{Bind}^{*}_{\mathcal{A}, \Pi}(\lambda)$, $\mathcal{A}$ observes a valid transaction with proof $\pi$ generated for an honest context $\Gamma^{*}_{hon}$. $\mathcal{A}$ then attempts to construct a valid transaction for a target context $\Gamma_{target}$ with proof $\pi^{\prime}$. $\mathcal{A}$ wins if $\text{VerifyAuth}$ outputs 1 and either the context deviates ($\Gamma_{target} \ne \Gamma_{hon}$) or the proof triggers a state transition that violates nullifier constraints. An SDAS is secure if $\mathcal{A}$ wins with probability $\le \text{negl}(\lambda)$.

\begin{figure*}[ht]
\centering
\resizebox{0.95\linewidth}{!}{%
\begin{tikzpicture}[
  >=latex',
  font=\small,
  signal/.style={->, thick},
  block/.style={draw, rounded corners, fill=gray!10, align=center,
                minimum height=2.8em, minimum width=4.8cm, inner sep=7pt},
  io/.style={draw, rounded corners, fill=blue!7, align=center,
             minimum height=3.0em, minimum width=7.4cm, inner sep=8pt},
  wit/.style={draw, rounded corners, fill=yellow!10, align=center,
              minimum height=3.0em, minimum width=7.4cm, inner sep=8pt},
  op/.style={draw, circle, fill=white, minimum size=1.7em, inner sep=0pt},
  note/.style={draw, rounded corners, fill=white, font=\footnotesize,
                align=left, inner sep=6pt},
  group/.style={draw, dashed, rounded corners, inner sep=10pt},
  lab/.style={fill=white, fill opacity=1, text opacity=1, draw=none, inner sep=2pt}
]

\pgfdeclarelayer{labels}
\pgfsetlayers{main,labels}

\node[io]  (PUB) at (4.2, 6.6) {\textbf{Public instance} $\mathbf{x}$\\
$x_{year},\ x_{thresh},\ x_{addr}$};

\node[wit] (WIT) at (4.2, 3.0) {\textbf{Private witness} $\mathbf{w}$\\
$w_{birth},\ w_{salt}$};

\begin{pgfonlayer}{labels}
\node[lab] (TY)    at (0.0, 4.7) {$x_{year}$};
\node[lab] (TB)    at (0.0, 2.0) {$w_{birth}$};
\node[lab] (SALT)   at (0.0, 0.5) {$w_{salt}$};
\node[lab] (CADDR) at (7.5, 6.6) {$x_{addr}$};
\end{pgfonlayer}

\node[op]    (SUB1)  at (10.0, 4.7) {$-$};
\node[block] (D1)    at (13.5, 4.7) {$age := x_{year}-w_{birth}$};

\node[block] (RANGE) at (20.0, 4.7) {Predicate Check\\$age \ge x_{thresh}$};
\node[block] (ACC)   at (26.0, 4.7) {Predicate\\accepted};

\node[block] (POSE)  at (16.0, 0.9) {Entropy Integration\\$saltSquare = w_{salt}^2$};

\node[block] (BIND)  at (16.0, 7.9) {Sender Binding\\$C_{addr} = x_{addr}^2$};

\draw[signal] (TY) -- (SUB1.west);
\draw[signal] (TB) |- (SUB1.south);

\draw[signal] (SUB1) -- (D1.west);
\draw[signal] (D1.east) -- (RANGE.west);
\draw[signal] (RANGE.east) -- (ACC.west);

\draw[signal] (SALT) |- ([xshift=-2pt]POSE.west);
\draw[signal] (CADDR) |- (BIND.west);

\draw[group] ($(PUB.north west)+(-0.5,0.5)$) rectangle ($(PUB.south east)+(0.5,-0.5)$);
\draw[group] ($(WIT.north west)+(-0.5,0.5)$) rectangle ($(WIT.south east)+(0.5,-0.5)$);

\draw[group] ($(SUB1.north west)+(-1.0,0.8)$) rectangle ($(ACC.south east)+(1.0,-1.25)$);
\node[anchor=south west, font=\footnotesize] at ($(SUB1.north west)+(-0.95,0.85)$) {\textbf{Age predicate gadget}};

\draw[group] ($(BIND.north west)+(-1.0,0.7)$) rectangle ($(BIND.south east)+(1.0,-0.7)$);
\node[anchor=south west, font=\footnotesize] at ($(BIND.north west)+(-0.95,0.75)$) {\textbf{Context binding gadget}};

\draw[group] ($(POSE.north west)+(-1.6,0.7)$) rectangle ($(POSE.south east)+(1.0,-0.7)$);
\node[anchor=south west, font=\footnotesize] at ($(POSE.north west)+(-0.95,0.75)$) {\textbf{Hiding commitment gadget}};

\end{tikzpicture}
}
\caption{Constraint layout of the \texttt{AgeCheck} circuit for age verification, sender binding, and salt-based witness expansion.\vspace{-3mm}}
\label{fig:circuit_logic}
\end{figure*}
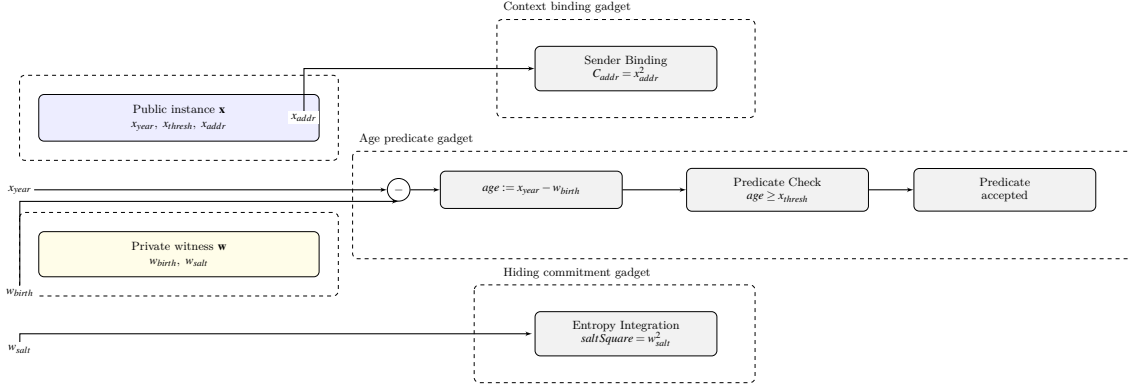

\section{\uppercase{ZK-Compliance}}
Our prototype uses Groth16 over the BN128 elliptic curve, which is supported by Ethereum precompiles for efficient on-chain verification. The implementation focuses on sender-bound authorization against mempool front-running. Cross-chain and cross-contract anchoring follow the same context-binding pattern, but remain implementation extensions of the current prototype.

\subsection{System Architecture}
ZK-Compliance uses a client-side proving model (Figure~\ref{fig:threat_model}). The raw identity attributes, such as birth year, remain in the local client environment and are never transmitted. This keeps private attributes outside the ledger’s persistent public state. Following the client-side workflow in Algorithm~\ref{alg:prover}, the client generates a proof locally using SnarkJS, binds the public input to the wallet address, and submits the signed \texttt{grantAccess} transaction to the registry contract. 

\begin{algorithm}[!ht]
 \caption{Client-Side Proof Generation Protocol}
 \label{alg:prover}
 \KwData{Birth year $w_{birth}$, Salt $w_{salt}$, Target Contract $addr_{SC}$}
 \KwResult{Signed Transaction $tx$}
 $x_{addr} \gets \text{Wallet.getAddress}()$\;
 $x_{year} \gets \text{System.getCurrentYear}()$\;
 $x_{thresh} \gets 18$\;
 $\mathbf{w} \gets \{ w_{birth}, w_{salt} \}$\;
 $\mathbf{x} \gets \{ x_{year}, x_{thresh}, x_{addr} \}$\;
 $pk \gets \text{FetchProvingKey}(\text{``circuit.zkey''})$\;
 $(\pi, \mathbf{x}) \gets \text{SnarkJS.fullProve}(\mathbf{w}, \mathbf{x}, pk)$\;
 $data \gets \text{ABI.encode}(\text{``grantAccess''}, \pi, \mathbf{x})$\;
 $tx \gets \text{Wallet.sign}(addr_{SC}, data)$\;
 \Return $tx$\;
\end{algorithm}

\subsection{Compliance Circuit Design}
We implement the \texttt{AgeCheck} circuit in Circom. Figure~\ref{fig:circuit_logic} shows its data flow and constraint layout. The circuit takes three public inputs, namely current year $x_{year}$, required age threshold $x_{thresh}$, and requesting address $x_{addr}$, and two private witnesses, birth year $w_{birth}$ and random salt $w_{salt}$. The circuit compiles to a 14-constraint Rank-1 Constraint System (R1CS) covering age checking, salt integration, and sender binding. It derives the user’s age as $\Delta = x_{year} - w_{birth}$ and verifies $\Delta \geq x_{thresh}$ without exposing $w_{birth}$. It integrates the private salt as $saltSquare = w_{salt}^2$, expanding the witness space against pre-computation. Finally, it constrains $C_{addr}=x_{addr}^2$ so the circuit consumes the public address input. Since $x_{addr}$ is in the proof’s public input vector and the contract checks $\mathbf{x}[2] == \texttt{msg.sender}$, changing the sender address invalidates verification.

\subsection{On-Chain Verification}
The verification contract acts as an access registry, mapping each user and auditor pair to an \texttt{AccessRecord} with an expiration timestamp and validity flag. When the contract receives a \texttt{grantAccess} transaction with proof $\pi$, public signals $\mathbf{x}$, duration $\delta$, and auditor $A$, it executes Algorithm~\ref{alg:verify}.

\begin{algorithm}[!ht]
 \caption{Prototype Verification with Front-Running Defense}
 \label{alg:verify}
 \KwData{Proof $\pi$, Public Signals $\mathbf{x}$, Duration $\delta$, Auditor $A$}
 \KwResult{On-chain Authorization State Update}
 \If{$\mathbf{x}[2] \neq \texttt{msg.sender}$}{
    \Return \textbf{Revert}(``Sender Mismatch'')\;
 }
 \If{$\mathbf{x}[0] \neq \texttt{CURRENT\_YEAR} \lor \mathbf{x}[1] \neq \texttt{REQUIRED\_AGE}$}{
    \Return \textbf{Revert}(``Invalid Protocol Constants'')\;
 }
 $valid \gets \texttt{Verifier.verifyProof}(\pi, \mathbf{x})$\;
 \If{$!valid$}{
    \Return \textbf{Revert}(``Invalid ZK Proof'')\;
 }
 $\sigma[\texttt{msg.sender}][A] \gets \{ \texttt{block.timestamp} + \delta, \text{TRUE} \}$\;
 \Return \textbf{Emit} \text{AccessGranted}(\texttt{msg.sender}, A)\;
\end{algorithm}

The first check enforces sender binding before cryptographic verification. If an adversary submits the proof from another wallet, \texttt{msg.sender} differs from the bound input $\mathbf{x}[2]$, causing an immediate revert. The contract then checks protocol constants, verifies the Groth16 proof using the BN128 precompile, and updates the \texttt{AccessRecord}. The update is idempotent, so replaying the same transaction overwrites the timestamp instead of creating redundant state.

\subsection{Revocation Mechanism}
\begin{figure}[ht]
\centering
\includegraphics[width=\linewidth]{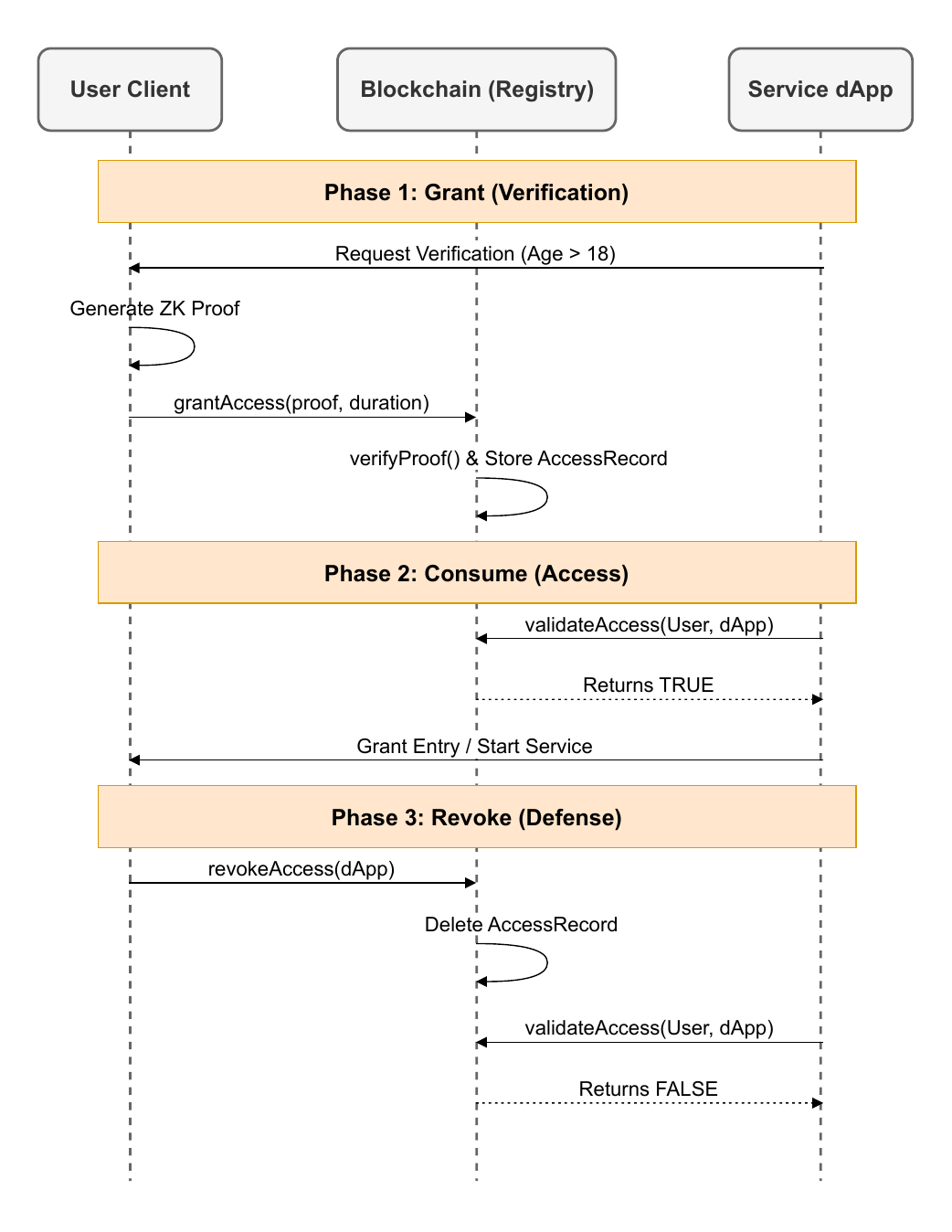}
\caption{Execution lifecycle.}
\label{fig:architecture_sequence_diagram}
\end{figure}

Users can sever access without contacting the original issuer by calling \texttt{revokeAccess()}, which deletes the corresponding \texttt{AccessRecord}. This reveals only that a pseudonymous address withdrew authorization for a dApp, not the underlying identity attributes. The current prototype deletes access state on revocation, but production deployments require stateful nullifiers to prevent stale proofs from reactivating access. The protocol lifecycle is summarized in Figure~\ref{fig:architecture_sequence_diagram}.

\begin{table}[t]
\centering
\caption{Prototype evaluation summary}
\label{tab:evaluation_summary}
\begin{tabular}{ll}
\toprule
Metric & Value \\
\midrule
Total constraints & 14 \\
Public inputs & 3 \\
Private inputs & 2 \\
Non-linear constraints & 8 \\
Witness generation & 42 ms \\
Proof construction & 142 ms \\
Total client latency & 184 ms \\
On-chain verification & 240,512 gas \\
\bottomrule
\end{tabular}
\end{table}

\section{\uppercase{Evaluation}}
\label{sec:evaluation}

We evaluate ZK-Compliance along three dimensions: circuit complexity, browser-based proving latency, and on-chain verification cost. The \texttt{AgeCheck} circuit was implemented in Circom 2.1 and compiled to a Rank-1 Constraint System. As shown in Table~\ref{tab:evaluation_summary}, the age predicate, entropy salt integration, and \texttt{msg.sender} binding require only $14$ constraints. This small circuit keeps proof generation feasible in browser-based dApp settings and suggests that sender binding adds minimal arithmetic overhead.

Client-side proving was measured using \texttt{snarkjs} compiled to WebAssembly in Chrome on a consumer-grade MacBook Air with an Apple M2 processor and 16GB RAM, simulating a standard web-based dApp interaction. The end-to-end proving process follows Algorithm~\ref{alg:prover}. Witness generation took $42$ ms and Groth16 proof construction took $142$ ms, giving a total client latency of $184$ ms. This sub-second latency supports interactive compliance checks without requiring external proof-generation services.

We deployed the \texttt{AgeGuardRegistry} contract to Sepolia and executed the \texttt{grantAccess} procedure in Algorithm~\ref{alg:verify}. The initial authorization grant cost $240{,}512$ gas, including sender anchoring, BN128 pairing verification, and storage of the \texttt{AccessRecord}. Because the prototype uses Groth16, the on-chain verification step remains constant-size with respect to the underlying identity predicate, giving users a predictable verification cost.

To validate sender binding, we simulated proof reuse by a different caller, where an adversarial wallet observed a valid \texttt{grantAccess} transaction and resubmitted the captured proof from another address. In all tests, the contract rejected the transaction because the bound public signal did not match \texttt{msg.sender}, preventing unauthorized transfer of access rights.

The cost trade-off is that ZK-Compliance stores authorization state on-chain. This increases gas consumption, but makes authorization explicit, auditable, and user-revocable. Unlike stateless proof presentation or off-chain session management, the smart contract enforces whether access remains valid. This design fits compliance settings where authorization must be verifiable, time-bounded, and withdrawable.

\section{\uppercase{Limitations}}
\label{sec:limitations}

The current prototype has two main limitations. First, our implementation evaluates sender binding against third-party mempool front-running, while cross-chain and cross-contract binding remain extensions of the SDAS context model. Future versions should add \texttt{block.chainid} and \texttt{address(this)} as public inputs to bind proofs to a specific chain and contract. Second, the revocation mechanism does not prevent self-replay of an unexpired proof after revocation. A production deployment should add stateful nullifiers, using a SNARK-friendly hash such as Poseidon~\cite{grassi2021poseidon}, to prevent stale proofs from reactivating revoked access.

\section{\uppercase{Conclusion}}
Public ledgers create tension between transparent verification and data minimization for regulatory compliance. We addressed this by formalizing SDAS, which separates private attribute validity from stateful on-chain authorization. Through ZK-Compliance, we implemented a ``Grant, Verify, Revoke'' lifecycle that binds a ZK proof to the executing \texttt{msg.sender}, preventing reuse by different callers while leaving chain, contract, and nullifier-based extensions for future work. Sepolia evaluation demonstrates the feasibility of sender-bound selective disclosure authorization on Ethereum, with browser-based proof generation under $200$ ms and on-chain verification costing $240,512$ gas. These results suggest that privacy-preserving compliance systems should treat authorization as a context-bound, revocable state transition instead of a stateless proof presentation.

\section*{\uppercase{Acknowledgements}}
We acknowledge the Data Agency and Security (DAS) Lab at George Mason University (GMU), and Google for partially supporting this work. The opinions expressed are solely those of the authors.







\bibliographystyle{apalike}
{\small
\bibliography{main}}



\end{document}